\documentclass{article}
\usepackage{spconf,amsmath,amssymb,graphicx,booktabs,xcolor,url}
\usepackage{cite}
\usepackage[inline]{enumitem}
\usepackage{tikz}
\usepackage{siunitx}
\usepackage{multirow}
\usetikzlibrary{arrows.meta,positioning,calc}
\graphicspath{{figures/}}

\title{Does per-frame early exit pay? A compute-matched study of dynamic depth for on-device speech enhancement}

\name{Cl\'ement Laroche$^{\star}$ \qquad Riccardo Miccini$^{\dagger}$%
\thanks{This project has received funding from the European Union's Horizon Europe research and innovation programme under the HORIZON-JU-Chips-2024-1-IA grant agreement No 101194172 (NeAIxt).}}
\address{$^{\star}$ GN A/S, Denmark \qquad $^{\dagger}$ Technical University of Denmark, Denmark}

\begin{document}
\ninept
\maketitle

\begin{abstract}
Deep learning-based speech enhancement is increasingly deployed on-device in hearing aids, headsets, and earbuds. Most of these devices, however, can only accelerate static \texttt{int8} graphs, so a depth-varying network must be implemented as several graphs, orchestrated by a policy. In this paper, we supervise every intermediate depth of one causal model, then we fine-tune its output heads to guarantee that deeper outputs are never worse than shallower ones. Using this training protocol, we can derive a family of static models that are more Pareto-efficient than their equivalently-sized counterparts trained from scratch on the same budget. Specifically, we achieve up to $0.11$ higher PESQ for equivalent compute, and match the best PESQ at \qty{30}{\percent} less compute. We then quantize the models to \texttt{int8} and measure the latency--quality frontier on an STM32N6 microcontroller. On VoiceBank--DEMAND, the dynamic enhancer lies on the same frontier as the static models, rather than trading quality for dynamic execution. Running the policy on the companion Cortex-M55 takes only \qty{26}{\micro\second} per frame, while splitting the enhancer into separate NPU graphs adds \qty{2.2}{\percent} latency overhead. The cost of dynamic execution is therefore small.

\end{abstract}

\section{Introduction}
\label{sec:intro}

Speech enhancement (SE) systems are a crucial part of telecommunications, teleconferencing, and assistive technology, improving remote collaboration, user experience, and quality of life.
In recent years, lightweight causal SE architectures have made significant progress, reaching competitive quality at just a few million multiply-accumulate operations (MACs)~\cite{braun2021nsnet2,rong2024gtcrn}.
However, deploying these systems on devices such as hearing aids, headsets, or earbuds requires a careful trade-off between denoising performance and computational needs, capable of satisfying real-time constraints within a shared battery budget.
Besides, device tiers and concurrent workloads ask for different quality/compute operating points; thus, a product needs a \textit{family} of models~\cite{cai2020ofa,shangguan2024todm}.
Because microcontroller-class accelerators support only static \texttt{int8} graphs, each member of that family is compiled ahead of time.
Consequently, a model that chooses its depth at run time is not something mainstream edge toolchains suport natively: they lower one static graph, leaving the selection policy and the dispatch between separately compiled subgraphs to run outside the compiled model~\cite{han2021dynamic,st_edgeai}.

Early-exiting dynamic neural networks place prediction heads along a backbone and route each input to one of them~\cite{han2021dynamic,kaya2019sdn}; within SE, multiple exits were introduced without a policy~\cite{miccini2023dynamic}, later using a probabilistic per-block criterion~\cite{olsen2026press}, while other dynamic approaches include learned width and gating~\cite{miccini2025scalable,miccini2025adaptive,zhao2026dynamically,parvathala2025gating}.
Train-once-deploy-many instead extracts many static sub-networks from one weight-shared run~\cite{huang2016stochastic,fan2020layerdrop,cai2020ofa,yu2019universally}, reportedly matching or surpassing dedicated training in on-device ASR~\cite{shangguan2024todm} and elastic music separation~\cite{li2024subnetwork}.
Several questions remain at the intersection of dynamic speech enhancement and embedded deployment. Dynamic SE is usually compared with its own exits~\cite{olsen2026press}, while retrained static models matched for compute are less often used as baselines~\cite{elminshawi2025dynslim}. Cost is also usually reported in nominal MACs, which ignore routing and dispatch overhead~\cite{ma2018shufflenetv2}. Finally, extracted static models have rarely been quantized and carried through to measured accelerator latency, and output quality has not been studied frame by frame. This leaves open whether the benefits of early-exit training come from adapting computation at runtime, from improving the static models extracted from the network, or from both.

In this paper, we add a per-frame \textit{monotonicity regularization} to a causal Conv-TasNet-style network, so that deeper exits are encouraged to improve over shallower ones. When training on the SE task loss alone, this is not guaranteed: on nearly half of the frames, a deeper exit performs worse than its previous one, confirming the ``overthinking'' phenomenon observed in early work on early-exiting~\cite{kaya2019sdn}. We address this by fine-tuning only the exit heads and penalizing such regressions during training, rather than correcting them after training as previously done for classification~\cite{jazbec2023anytime}.

Once deeper exits become consistently useful, the resulting weight-shared model can serve two roles: it can be routed dynamically, or split into a family of fixed-depth models. We compare both choices against recipe-matched retrained static models and quantize the models to \texttt{int8}, allowing them to run on the Neural-ART NPU integrated in an STM32N6 microcontroller, thereby measuring the true on-device latency. The main observation is that, once the fixed-depth models are compared fairly, routing brings little advantage: indeed, the extracted static models form a competitive quality--compute frontier, which holds when considering measured latency instead of theoretical MACs. Our claims are limited to speech-dense datasets (VoiceBank--DEMAND); while arguably more realistic, workloads with long pauses are outside this study.

We contribute the following:
\begin{enumerate*}[label={\arabic*)}]
\item A per-frame monotonicity regularization for regression exits, together with a heads-only fine-tuning strategy;
\item An analysis showing that much of the routing headroom comes from frames where deeper exits reduce quality, and that repairing these frames largely removes this advantage;
\item A compute-matched static frontier retrained under the same recipe and evaluation harness, with seed spreads and controls;
\item \texttt{int8} quantization and measured STM32N6 latency for every extracted fixed-depth model.
\end{enumerate*}

\section{Early-exit Conv-FSENet}
\label{sec:setup}

\begin{figure}[t]\centering
\begin{tikzpicture}[
  x=1mm, y=1mm, >={Latex[length=1.3mm]},
  blk/.style={draw, rounded corners=1pt, minimum height=5mm, minimum width=8.5mm,
              inner sep=0.5pt, align=center, font=\scriptsize},
  hd/.style ={draw, rounded corners=1pt, minimum height=4mm, minimum width=8.5mm,
              inner sep=0.5pt, align=center, font=\scriptsize, fill=black!7},
  rt/.style ={draw, dashed, rounded corners=1pt, minimum height=4mm,
              inner sep=1.5pt, align=center, font=\scriptsize},
  n/.style  ={font=\scriptsize, inner sep=0.5pt},
  t/.style  ={font=\tiny, inner sep=0.5pt},
]
\def\dx{17}
\node[blk] (fe) at (0,0)        {$1{\times}1$};
\node[blk] (s1) at (\dx,0)      {stack 1};
\node[blk] (s2) at (2*\dx,0)    {stack 2};
\node[blk] (s3) at (3*\dx,0)    {stack 3};
\node[n, left=5mm of fe] (in) {$|X|^{0.3}$};
\draw[->] (in) -- (fe);
\draw[->] (fe) -- (s1) coordinate[midway] (mw);
\draw[->] (s1) -- (s2);
\draw[->] (s2) -- (s3);
\node[t, above=0.3mm of fe] {$257{\to}128$};
\node[t, above=0.3mm of s1] {dil.\ 1/2/4};
\node[t, above=0.3mm of s2] {1/2/4};
\node[t, above=0.3mm of s3] {1/2/4};
\foreach \k in {0,1,2,3} {
  \pgfmathsetmacro\xx{\k*\dx}
  \node[hd] (h\k) at (\xx,-9) {head \k};
}
\foreach \a/\b in {fe/h0, s1/h1, s2/h2, s3/h3} \draw[->] (\a) -- (\b);
\node[rt] (rt) at (1.5*\dx,-17) {one-shot router (GRU-16)};
\draw[->] (mw) |- (rt.west);
\end{tikzpicture}
\caption{The exit ladder. Head~0 reads the frontend alone (the \emph{bypass}
exit); head~$i$ reads that plus stacks $1..i$. The one-shot router feeds on frontend features before any stack runs.}
\label{fig:ladder}
\end{figure}
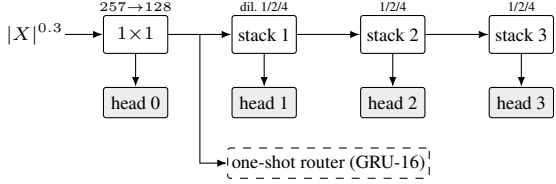

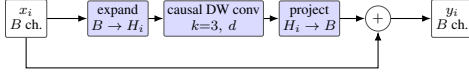
\begin{figure}[t]
\centering
\resizebox{0.72\linewidth}{!}{%
\begin{tikzpicture}[
  font=\footnotesize,
  blk/.style={draw=black!70, rounded corners=1pt, minimum height=7.5mm,
              align=center, inner sep=2.5pt, fill=white},
  stack/.style={draw=black!70, rounded corners=2pt, inner sep=4pt,
                fill=blue!8},
  inner/.style={draw=black!70, rounded corners=1pt, minimum height=7mm,
                align=center, inner sep=2pt, fill=blue!14},
  add/.style={draw=black!70, circle, inner sep=1pt, minimum size=5mm,
              fill=white},
  arr/.style={-{Latex[length=1.8mm]}, semithick}
]

\node[blk] (xin) at (0,0) {$x_i$\\$B$ ch.};

\node[inner, right=8mm of xin] (exp) {expand\\$B \to H_i$};
\node[inner, right=3mm of exp] (dw)
  {causal DW conv\\$k{=}3,\ d$};
\node[inner, right=3mm of dw] (proj)
  {project\\$H_i \to B$};
\node[add, right=5mm of proj] (add) {$+$};

\node[blk, right=8mm of add] (yout) {$y_i$\\$B$ ch.};

\draw[arr] (xin) -- (exp);
\draw[arr] (exp) -- (dw);
\draw[arr] (dw) -- (proj);
\draw[arr] (proj) -- (add);
\draw[arr] (add) -- (yout);

\draw[arr]
  (xin.south) -- ++(0,-6mm)
  -| (add.south);

\end{tikzpicture}}
\caption{One residual block $B_i$ within stack $S_i$.}
\label{fig:stack_simple}
\end{figure}

\subsection{Architecture and exits}
\label{sec:architecture}

Given a single-channel noisy signal $x$, composed of clean speech $s$ and additive noise $n$, we estimate the speech as $\hat{s}=\mathcal{M}(x)$ using Conv-FSENet~\cite{miccini2025scalable,miccini2025diet}, a causal STFT-domain masking network derived from Conv-TasNet \cite{luo2019convtasnet}. The model comprises a pointwise-convolutional frontend, $R$ stacks, each featuring 3 causal dilated depthwise temporal-convolutional residual blocks (shown in Fig.~\ref{fig:stack_simple}), and mask heads.

The frontend maps the compressed noisy magnitude $|X|^{0.3}$ onto a $B$-channel inter-block representation, which is expanded into $H_i$ channels within each $i$-th block before projecting back to $B$.

To reduce MACs, we allow the network to stop at different depths instead of evaluating all $R$ TCN stacks on every frame. Mask heads attach after the frontend (the \emph{bypass} exit) and at each stack, estimating a sigmoid-activated magnitude mask on the noisy complex STFT, such that exit $i$ evaluates the frontend plus the first $i$ stacks (Fig.~\ref{fig:ladder}). Their receptive fields are $1$, $15$, $29$ and $43$ frames, while the bypass exit is memoryless.
At each \qty{16}{\milli\second} frame, a one-shot policy router selects an exit before executing any of the stacks; this consists of a 16-unit GRU, feeding on the frontend output and producing a categorical decision $k(t)\in\{0,\ldots,R\}$ before any TCN stack is executed. Only stacks up to $k(t)$ are then evaluated. The remaining stacks are skipped entirely. The resulting compute for a frame is therefore approximately
\begin{equation}
C_{k(t)}
=
C_{\mathrm{fe}}
+
\sum_{i=1}^{k(t)} C_{\mathrm{stack},i}
+
C_{\mathrm{head},k(t)},
\end{equation}
where
\begin{equation}
C_{\mathrm{fe}} \approx FB,
\quad
C_{\mathrm{stack},i}
\approx
3H_i(2B+3),
\quad
C_{\mathrm{head},k(t)} \approx BF.
\end{equation}

Selecting an earlier exit therefore avoids the full cost of every subsequent TCN stack, while the frontend and selected mask-head costs remain fixed.

\subsection{Training and exit policy}
\label{sec:recipe}

Our enhancement loss $\mathcal{L}$ is the power-law-compressed spectral MSE of As in \cite{braun2021loss,miccini2023dynamic}, computed on active-speech-level-normalized spectra. Magnitude and complex terms use a compression exponent $c{=}0.3$ and are mixed with weight $\alpha{=}0.3$.

\begin{equation}
\label{eq:loss_se}
\begin{aligned}
\mathcal{L}_{\mathrm{SE}}(\widehat{S},S)
={}&
\alpha
\sum_{l,f}
\left|
|S|^c e^{j\angle S}
-
|\widehat{S}|^c e^{j\angle \widehat{S}}
\right|^2
\\
&+
(1-\alpha)
\sum_{l,f}
\left(
|S|^c-|\widehat{S}|^c
\right)^2 .
\end{aligned}
\end{equation}
where $l$ and $f$ denote time and frequency bins. We use $c{=}0.3$ and $\alpha{=}0.3$, and compute the loss on active-speech-level-normalized spectra.

For the early-exit model, $\mathcal{L}_{\mathrm{SE}}$ is applied to the output selected by the router and, under deep supervision (DS) \cite{lee2015dsn,llombart2019progressive}, independently to every exit. The enhancement part of the objective is therefore
\begin{equation}
\mathcal{L}_{\mathrm{enh}}
=
\mathcal{L}_{\mathrm{SE}}(\widehat{S}_{k},S)
+
\frac{\lambda_{\mathrm{ds}}}{R+1}
\sum_{j=0}^{R}
\mathcal{L}_{\mathrm{SE}}(\widehat{S}_{j},S),
\label{eq:loss_enh}
\end{equation}
where $\widehat{S}_{k}$ is the routed output and $\widehat{S}_{j}$ is the output of exit $j$.

The policy is supervised with a clean-signal \emph{knee target}, which identifies the point of diminishing returns along the exit ladder. Using the same compression exponent, we define the per-frame compressed-magnitude error of exit $j$ as
$$
q_j(t)=\frac{1}{F}\sum_f\left(|\hat S_j(f,t)|^c-|S(f,t)|^c\right)^2.
$$
The knee target then selects the cheapest exit whose error is within $\varepsilon$ of the best achieved by any exit:
\begin{equation}
k(t)=\min\left\{
j : q_j(t)\leq \min_i q_i(t)+\varepsilon
\right\}.
\label{eq:knee}
\end{equation}

We train the policy by cross-entropy against this target, either jointly with the full backbone or on a frozen enhancer. Routing is discretized by straight-through Gumbel-softmax~\cite{jang2017gumbel}.

Combining enhancement, exit supervision, routing, and compute control gives
\begin{equation}
\mathcal{L}_{\mathrm{tot}}
=
\mathcal{L}_{\mathrm{enh}}
+
\lambda_{\mathrm{knee}}
\mathcal{L}_{\mathrm{CE}}
+
\mathcal{L}_{\mathrm{cost}} .
\label{eq:loss_total}
\end{equation}
We use $\lambda_{\mathrm{ds}}{=}1$ and $\lambda_{\mathrm{knee}}{=}0.1$, while $\mathcal{L}_{\mathrm{cost}}$ controls the average routed compute through the target operating point $\tau$.

\subsection{Repairing the exits}

Because the knee target is defined from the current exits, it does not require quality to improve with depth. In practice, a deeper exit can perform worse than the previous one (Sec.~\ref{sec:recipe}). This creates an ambiguity when evaluating routing: a policy may appear useful simply because it avoids harmful deeper exits, rather than because it assigns less compute to frames that genuinely need less processing. We therefore repair the exit ladder before assessing the value of dynamic depth.

To separate these effects, we apply a \emph{monotonicity repair} to the exits themselves. We freeze the backbone stacks, frontend, and policy, and fine-tune only the exit heads ($4{\times}33$k parameters) with deep supervision and the following regularizer:
\begin{equation}
\mathcal{L}_{\mathrm{mono}}=
\frac{1}{TR}\sum_{t=1}^{T}\sum_{j=0}^{R-1}
\max\bigl(0,\;q_{j+1}(t)-q_j(t)\bigr),
\label{eq:mono}
\end{equation}
which penalizes every frame for which adding depth increases the error. We fine-tune for \num{2000} steps with the Adam optimizer with a learning rate of $10^{-4}$ and a regularization weight of 50. Rather than enforcing monotonicity post hoc, as Jazbec et al.~\cite{jazbec2023anytime} do for classification, this term encourages it through a soft constraint during training.

\section{Results}
\label{sec:results}

\subsection{Experimental setup}

\begin{description}[
    font=\bfseries,
    leftmargin=0pt,
    parsep=\parsep,
    listparindent=\parindent,
    labelwidth=0em,
    itemindent=1em,
    labelsep=1em,
    align=left,
    itemsep=\parsep,
]
\item[Data]
We use the 16-kHz VoiceBank--DEMAND dataset \cite{valentini2016vbd}. During training, we randomly crop the audio into \qty{4}{\second} segments and apply Remix/BandMask/Shift augmentation \cite{defossez2020demucs}, together with the spectral augmentation and level-invariance techniques from \cite{Braun_2020}. We compute the STFT using a \num{512}-sample window with \qty{50}{\percent} overlap, yielding $F{=}257$ bins, and use $|X|^{0.3}$ as input to all models.

\item[Models and static baselines] 
As a baseline for per-frame routing, we compare it with static models at the same compute budget. We train a grid of Conv-FSENets varying TCM width $H\in{32,64,96,256}$ and number of stacks $R\in{1,2,3}$, while keeping $B{=}128$ fixed. The statics use the same backbone with the routing and multi-exit machinery removed. The twelve models span $5.7$--$41.8$\,MMAC/s and bracket all routed points ($6.6$--$20.2$\,MMAC/s). Three exactly cost-matched width--depth pairs further show whether the same compute budget is better spent on width or depth.

\item[Training] 
All models use the same enhancement objective and are trained with Adam ($10^{-3}$ learning rate, $10^{-5}$ weight decay, and $0.99$/epoch decay), batch size 64, and early stopping. The static models follow the same training protocol, with routing-specific losses removed.

\item[Metrics] 
We report whole-clip SI-SDR~\cite{leroux2019sisdr} and wideband PESQ~\cite{itu2001pesq} on all 824 test clips. Dynamic compute is the realized average MMAC/s over the selected exits.%
\end{description}

\subsection{Quality--compute frontier}

\begin{figure}[t]\centering
\includegraphics[width=0.95\linewidth]{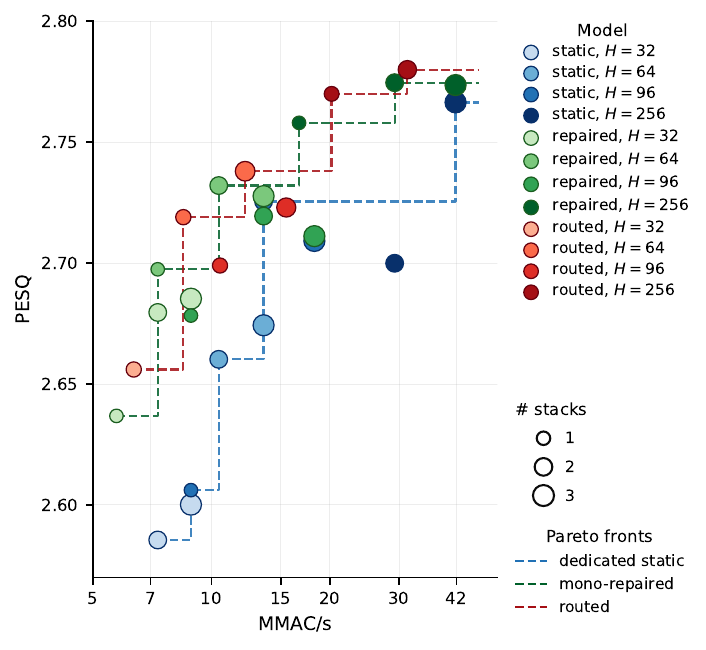}
\caption{Pareto front for \texttt{Float32} deployments on the test set.}
\label{fig:pareto}
\end{figure}

Figure~\ref{fig:pareto} compares the quality--compute trade-off of the dedicated static baselines, the routed models, and the mono-repaired exits. Against the dedicated statics, routing gives a clear advantage over most of the compute range. For example, the routed $H{=}64$ model reaches $2.719$ PESQ at $8.48$\,MMAC/s, while the static frontier is around $2.60$ at the same cost. The gap becomes smaller as compute increases, but the routed models remain above the dedicated-static frontier.

The picture changes once each exit is repaired as a static model. The mono-repaired frontier almost overlaps the routed one: at matched compute, their PESQ differs by less than $0.01$ over the common range. For instance, the routed points at $8.48$, $12.18$, and $20.22$\,MMAC/s differ from the interpolated repaired frontier by only $+0.009$, $-0.001$, and $+0.007$ PESQ, respectively. 

\begin{figure}[t]\centering
\includegraphics[width=\linewidth]{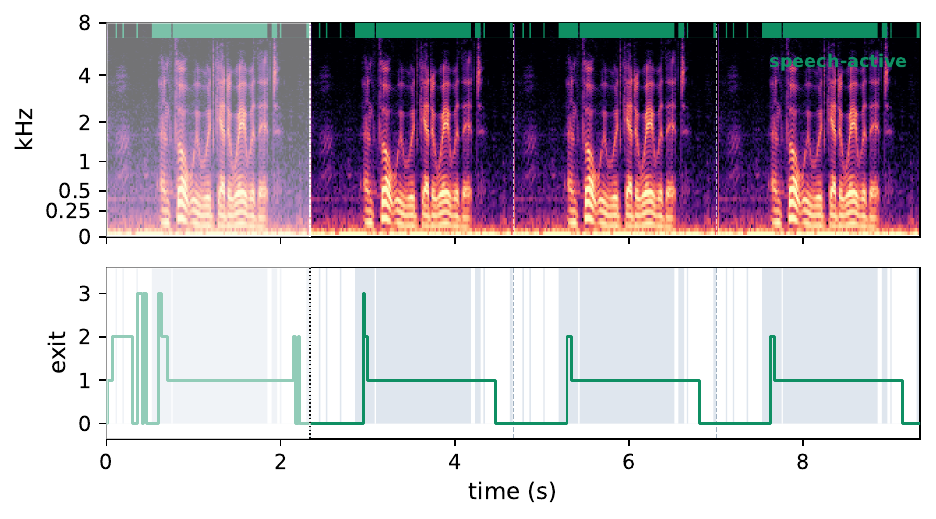}
\caption{Example router behaviour on a difficult VoiceBank--DEMAND utterance (0.46,dB input SNR, noisiest \qty{5.6}{\percent} of the test set, \qty{65.5}{\percent} speech-active), repeated four times; the faded first repeat warms the GRU and is discarded. The router spends depth mainly during speech: \qty{74}{\percent} of speech-inactive frames take the bypass, versus \qty{12}{\percent} of speech frames, reducing compute by \qty{58.7}{\percent} over the warmed repeats relative to always running the deepest exit.}

\label{fig:router_timeline}
\end{figure}

The similar routed and repaired frontiers show that per-frame routing does not raise the best quality attainable at a given compute budget, but importantly it does not noticeably lower it either. This makes the deployment choice workload-dependent. A repaired static model fixes one operating point and pays the same cost on every frame, whereas the routed model keeps the full exit scaffold and can move between operating points over time. Figure~\ref{fig:router_timeline} illustrates this behaviour on a repeated VoiceBank--DEMAND utterance: during speech, the router mostly selects a processing exit, while in speech-inactive regions it falls back to the cheap bypass. Across the test set, bypass selection is $2.4$--$3.4\times$ more frequent in silence than in speech. On dense workloads with little variation, a static deployment is therefore sufficient; on streams with pauses or changing difficulty, routing can reduce average compute while remaining on essentially the same quality--compute frontier. The benefit of dynamic deployment is thus not a higher quality ceiling, but the ability to adapt where compute is spent without sacrificing that ceiling.

\section{Measured Latency on a microcontroller}

\subsection{deployment platform}

We evaluate deployment on the STM32N6570-DK, a microcontroller platform combining an Arm Cortex-M55 host processor with STMicroelectronics' Neural-ART neural accelerator. The NPU targets statically shaped, integer-quantized workloads, while unsupported operations and graph orchestration remain on the Cortex-M55. All networks are causal at a \qty{16}{\milli\second} hop. We keep the STFT, iSTFT and masking on the host (no NPU FFT) and export ring buffers as explicit state tensors, making each graph a pure per-frame stateless function. The models are exported to ONNX and post-training \texttt{int8} quantized~\cite{nagel2021quant} with per-channel scaling and entropy calibration, then compiled through ST Edge AI Core v4.0~\cite{st_edgeai} with operator rewrites where needed. Compiler reports provide the NPU/host partition and memory requirements. For the experiments below, Neural-ART runs at \qty{1}{\giga\hertz} and the Cortex-M55 at \qty{800}{\mega\hertz}.

\subsection{MACs versus measured latency}
Hardware latency depends not only on the number of operations, but also on how they are arranged. Table~\ref{tab:macvsms} shows this clearly on the STM32N6: across ten measured topologies, execution time per analytic MMAC varies by $1.7\times$, from $82$ to $137,\mu$s/MMAC. Depth is particularly costly because each additional stack requires another accelerator launch, while increasing width makes better use of the NPU. For example, increasing $H$ from $32$ to $256$ raises the MACs within a stack by $8\times$, but its measured latency by only $3.1\times$. Consequently, models with the same nominal cost can have different runtimes: $(H{=}32,R{=}2)$ and $(H{=}64,R{=}1)$ have the same MACs and parameter count, yet take $1.00$ and $0.75$,ms per frame, respectively. MAC count can even give the wrong ordering: $(H{=}96,R{=}1)$ uses \qty{21}{\percent} more MACs than $(H{=}32,R{=}2)$ but runs \qty{15}{\percent} faster. We therefore do not use nominal MACs to estimate deployment cost, and instead measure every static, repaired, and routed workload directly on the STM32N6.

\begin{table}[t]
\centering
\caption{Analytic MACs do not price graphs on the STM32N6 NPU. Ten complete
int8 single-graph measurements (same board and toolchain).}
\label{tab:macvsms}
\scriptsize
\setlength{\tabcolsep}{11pt}
\begin{tabular}{lccc}
\toprule
Topology & MMAC/s & ms/frame & $\mu$s/MMAC \\
\midrule
$H{=}32$, $R{=}1$  &  5.73 & 0.640 & 112 \\
$H{=}64$, $R{=}1$  &  7.30 & 0.751 & 103 \\
$H{=}32$, $R{=}2$  &  7.30 & 1.001 & \textbf{137} \\
$H{=}96$, $R{=}1$  &  8.86 & 0.852 & 96 \\
$H{=}64$, $R{=}2$  & 10.43 & 1.215 & 117 \\
$H{=}96$, $R{=}2$  & 13.57 & 1.427 & 105 \\
$H{=}256$, $R{=}1$ & 16.70 & 1.364 & \textbf{82} \\
$H{=}96$, $R{=}3$  & 18.27 & 2.078 & 114 \\
$H{=}256$, $R{=}2$ & 29.25 & 2.472 & 85 \\
$H{=}256$, $R{=}3$ & 41.79 & 3.584 & 86 \\
\bottomrule
\end{tabular}
\end{table}

\subsection{Dynamic execution}

\begin{description}[
    font=\bfseries,
    leftmargin=0pt,
    parsep=\parsep,
    listparindent=\parindent,
    labelwidth=0em,
    itemindent=1em,
    labelsep=1em,
    align=left,
    itemsep=\parsep,
]
\item[Model export] 
As illustrated in Fig.~\ref{fig:ladder}, the dynamic models are deployed as nine subgraphs for the frontend, router, three stacks, and four exit heads, with the routing decision and dispatch handled by the host. Because different depths share the same frontend and stack graphs, the full family requires the same storage of a single ladder rather than several separate models. Quantized outputs remain close to float with the shared compressed-magnitude $|X|^{0.3}$ input: the repaired family changes by only $-0.03$ to $+0.01$ PESQ, and quantization shifts routed compute by at most \qty{2}{\percent}. We report measured per-frame mean latency on the STM32N6570; since the policy router runs on the dedicated M55 core, we report its contribution from measured cycle counts at its \qty{800}{\mega\hertz} clock.

\item[State under skipping] 

Each stack maintains its own ring buffers for the dilated convolutions. These buffers are updated only when that stack runs, so when a skipped stack resumes, its state reflects the most recent frames processed by that stack rather than the most recent input frames.
This is far more realistic than common offline evaluation protocols, which compute the routed output as a gated sum, thereby effectively updating each exit even when unused. 
Driving the exported graphs on the deployed path costs $-0.005{\pm}0.011$ PESQ over 824 clips and $0.14$\,dB SI-SDR, with routing decisions bit-identical across arms. Neither zero-order hold nor propagating the last computed state to deeper stacks \cite{elbayad2020depth} recovers the lost quality. We therefore keep skipped-stack states unchanged, which is already the behavior of the exported graphs and requires no firmware modification.

\end{description}

\subsection{On-Device results}

Table \ref{tab:n6_dynamic} shows the deployed latency for a representative set of repaired static and routed models. Within a fixed ladder, early exit provides a clear reduction in average execution time. For example, repaired routed $H{=}64$ at $\varepsilon{=}2{\cdot}10^{-3}$ runs at $0.930$,ms/frame, compared with $1.215$\,ms for the corresponding $H{=}64,R{=}2$ static, while PESQ changes from $2.716$ to $2.694$. At $H{=}256$, routing reduces latency from $2.472$ to $1.668$,ms/frame while retaining $2.773$ versus $2.781$ PESQ. Thus, when the backbone is fixed, routing can avoid substantial work with little quality loss.

The advantage is much smaller once the static topology is also allowed to change. Several routed points are matched or dominated by a different static width/depth choice: routed $H{=}32$ gives $2.639$ PESQ at $0.769$,ms/frame, while static $H{=}64,R{=}1$ reaches $2.684$ at $0.751$,ms; routed $H{=}96$ similarly reaches $2.683$ at $1.040$,ms. The more quality-oriented routed configurations show the same trend: routed $H{=}64$ with asym4 reaches $2.724$ at $1.473$,ms, whereas static $H{=}256,R{=}1$ reaches $2.762$ at $1.364$,ms. Dynamic depth therefore mainly provides intermediate operating points within one deployed ladder; it does not create a better latency--quality frontier than architecture-matched statics.

\begin{table}[t]
\centering
\caption{Measured STM32N6 latency for representative repaired static and routed models.}
\label{tab:n6_dynamic}
\scriptsize
\setlength{\tabcolsep}{4.0pt}
\begin{tabular}{llccc}
\toprule
Model & Variant & MMAC/s & int8 PESQ & ms/frame \\
\midrule
\multicolumn{5}{l}{\emph{Within a fixed ladder}} \\
Static $H{=}64,R{=}2$      & fixed              & 10.4 & 2.716 & 1.215 \\
Routed $H{=}64$            & $\varepsilon{=}2{\cdot}10^{-3}$ & 8.58 & 2.694 & \textbf{0.930} \\
\addlinespace[1pt]
Static $H{=}256,R{=}2$     & fixed              & 29.3 & 2.781 & 2.472 \\
Routed $H{=}256$           & $\varepsilon{=}2{\cdot}10^{-3}$ & 20.19 & 2.773 & \textbf{1.668} \\
\midrule
\multicolumn{5}{l}{\emph{Across static and dynamic topologies}} \\
Static $H{=}64,R{=}1$      & fixed              & 7.3  & \textbf{2.684} & \textbf{0.751} \\
Routed $H{=}32$            & $\varepsilon{=}2{\cdot}10^{-3}$ & 6.34 & 2.639 & 0.769 \\
Routed $H{=}96$            & $\varepsilon{=}2{\cdot}10^{-3}$ & 10.51 & 2.683 & 1.040 \\
\addlinespace[1pt]
Static $H{=}256,R{=}1$     & fixed              & 16.7 & \textbf{2.762} & \textbf{1.364} \\
Routed $H{=}64$            & $\varepsilon{=}5{\cdot}10^{-4}$ & 12.29 & 2.724 & 1.473 \\
\bottomrule
\end{tabular}
\end{table}

\section{Conclusion}
\label{sec:conclusion}
We presented a training recipe for early-exiting SE models that produces a family of deployable \texttt{int8} graphs from a single run, while retaining the option of dynamic routing. 
Through the aforementioned regularization strategy, we promote monotonically-increasing performance across the exit latter, thereby achieving \numrange{0.04}{0.11} PESQ improvements under the same training budget.
We could also match the best model (in terms of PESQ) with \qty{30}{\percent} less compute, although at a \qty{1.0}{\decibel} SI-SDR drop.
We observe how these gains are robust to quantization, and when deployed to the STM32N6 unit, constitute a latency--quality front that subsequent routed models can also match. On speech-dense datasets, however, per-frame depth selection brings little additional benefit. 
The apparent routing opportunity largely come from frames where deeper exits degrade the result, and the monotonicity repair removes much of that signal. The early-exit ladder can therefore serve both as useful training scaffolding for train-once deploy-anywhere networks as well as deployable dynamic models that can be operated without end-to-end routing. 
Whether automatic routing becomes more valuable on sparser workloads, where long pauses and cheap bypass opportunities are common, remains an important next step.

\bibliographystyle{IEEEbib}
\bibliography{refs}

\end{document}